\documentclass{ws-ijgmmp}
\def\be{\begin{equation}}
\def\ee{\end{equation}}
\def\ba{\begin{eqnarray}}
\def\ea{\end{eqnarray}}

\def\be{\begin{equation}}
\def\ee{\end{equation}}
\def\bea{\begin{eqnarray}}
\def\eea{\end{eqnarray}}

\def\yzero{\smash{\hbox{$y\kern-4pt\raise1pt\hbox{${}^\circ$}$}}}

\def\m{\mu}

\def\beq{\begin{equation}}
\def\eeq{\end{equation}}
\def\beqa{\begin{eqnarray}}
\def\eeqa{\end{eqnarray}}

\def\-{\hphantom{-}}

\def\s2{\frac{1}{\sqrt2}}

\def\beq{\begin{equation}}
\def\eeq{\end{equation}}
\def\beqa{\begin{eqnarray}}
\def\eeqa{\end{eqnarray}}

\def\IF{\relax{\rm I\kern-.18em F}}
\def\II{\relax{\rm I\kern-.18em I}}
\def\IP{\relax{\rm I\kern-.18em P}}
\def\IC{\relax\hbox{\kern.25em$\inbar\kern-.3em{\rm C}$}}
\def\IR{\relax{\rm I\kern-.18em R}}

\def\Dsl{\,\raise.15ex\hbox{/}\mkern-13.5mu D} 
\def\IZ{Z\kern-.4em  Z}

\begin{document}

\markboth{Authors' Names}
{Instructions for Typing Manuscripts (Paper's Title)}

%
\catchline{}{}{}{}{}

\title{T-duality Invariance of the Supermembrane}
\author{Mar\'\i a Pilar Garc\'\i a del Moral}
\address{Departamento de F\'\i sica, Universidad de Oviedo, Avda Calvo
Sotelo S/n. Oviedo,33007  Espa\~na\}\\
\email{garciamormaria@uniovi.es} }
\author{Joselen Pe\~na }
\address{Departamento de F\'\i sica, Facultad de Ciencias,\\
 Universidad Central de Venezuela,
 A.P. 47270, Caracas 1041-A, Venezuela\\
jpena@ciens.fisica.ucv.ve}
\author{Alvaro Restuccia}
\address{Departamento de F\'\i sica, Universidad de Antofagasta, Aptdo 02800, Chile \\
 $\&$ Departamento de F\'\i sica, Universidad Sim\'on Bol\'\i var\\
Apartado 89000, Caracas 1080-A, Venezuela\\
arestu@usb.ve }
\maketitle
\begin{history}
\received{(Day Month Year)}
\revised{(Day Month Year)}
\end{history}

\begin{abstract}We show that the supermembrane theory compactified on a torus is invariant under T-duality. There are two different topological sectors of the compactified supermembrane (M2) classified according to a vanishing or  nonvanishing second cohomology class. We find the explicit T-duality transformation that acts locally on the supermembrane theory and we show that it is an exact symmetry of the theory. We give a global interpretation of the T-duality in terms of bundles. It has a natural description in terms of the cohomology of the base manifold and the homology of the target torus. We show that in the limit when the torus degenerate into a circle and the M2 mass operator restricts to the string-like configurations, the usual closed string T-duality transformation between the type IIA and type IIB mass operators is recovered. Moreover if we just restrict M2 mass operator to string-like configurations but we perform a generalized T-duality we find the SL(2,Z) non-perturbative multiplet of IIA.
\end{abstract}


\section{Introduction}
In this note we show that the supermembrane compactified on a torus is invariant under a generalized T-duality transformation \cite{gmpr2}. As a new result, we show how in the String Theory limit, the T-duality transformation for the supermembrane becomes the standard T-duality transformation of the closed superstrings compactified on a circle.
The compactified supermembrane on a target space $M_9\times T^2$ may be formulated in terms of sections on symplectic torus bundles \cite{gmmpr}. There are two well-defined sectors: one in which a topological condition due to an irreducible wrapping is imposed, that corresponds to the so-called supermembrane with central charges \cite{mor}, and a second one without that restriction. While the first one can be globally formulated in terms of sections on symplectic torus bundles with $SL(2,\mathbb{Z})$ monodromy, the second corresponds to the formulation on trivial symplectic torus  bundles \cite{gmpr2}. Physically the two sectors have very different properties, among which the most relevant one is that the regularized supermembrane with central charges has discrete spectrum \cite{bgmr} -so it is a well-defined quantum object- in distinction with the compactified supermembrane without that condition \cite{dwpp}.

 In \cite{4hull04} it was argued that a fundamental
formulation of string/M-theory should exist in which the T- and
U-duality symmetries are manifest from the start.  In particular, it
was argued that many massive, gauged supergravities cannot be
naturally embedded in string theory without such a framework
\cite{stw},\cite{6hullreid},\cite{samtleben}. In \cite{gmpr2} we showed that the supermembrane on a torus with central charges  is in fact the origin of type IIB, and IIA gauged supergravities in 9D. 
We  showed the existence of a new $Z_2$ symmetry
that plays the role of T-duality in the supermembrane interchanging the
winding and KK charges but leaving the Hamiltonian invariant. T-duality becomes an exact symmetry of the
symplectic torus bundle description of the supermembrane by fixing its energy tension.
\section{The Compactified Supermembrane in $M_9\times T^2$}
We consider now the compactified supermembrane embedded on a target space $M_9\times T^2$  where
$T^2$ is a flat torus. We consider
maps $X^m,X^r$  from $M_9\times T^2$ to the target space , where $X^m$ with $m=3,\dots,9$ are single valued maps onto the Minkowski sector of the target
space while $X^r$, with $r=1,2$ maps onto the $T^2$ compact sector of the
target. The winding condition corresponds to
\bea\begin{aligned}
\oint_{\mathcal{C}_{s}}dX^1=2\pi R (l_{s}+m_{s}Re\tau);\quad
\oint_{\mathcal{C}_{s}}dX^2=2\pi R m_{s}Im\tau;\quad
\oint_{\mathcal{C}_{s}}dX^m=0 \end{aligned}\eea where  $R,\tau$ are respectively the radius and the Teichmuller parameters of the 2-torus target-space, and $l_{s},m_{s}, s=1,2$, are integers.
The physical hamiltonian in the LCG is given by \begin{equation}\label{5}
\begin{aligned}
&\mathcal{H}= \int_{\Sigma}T_{11}^{-2/3}\sqrt{W}\left[\frac{1}{2}(\frac{P_{m}}{\sqrt{W}})^{2}+
\frac{1}{2}(\frac{P_{r}}{\sqrt{W}})^{2}
+\frac{T_{11}^{2}}{2}\{X^{r},X^{m}\}^{2}
+\frac{T_{11}^{2}}{4}\{X^{r},X^{s}\}^{2}\right]\\ \nonumber &+\int_{\Sigma}T_{11}^{-2/3}\sqrt{W}\left[\frac{T_{11}^{2}}{4}\{X^{m},X^{n}\}^{2}- \overline{\Psi}\Gamma_{-} \Gamma_{m} \{X^m,\Psi\}- \overline{\Psi}\Gamma_{-} \Gamma_{r} \{X^r,\Psi\}\right] \end{aligned}\end{equation} subject to the constraints 
\begin{equation}  \begin{aligned}\label{e2}
\phi_{1}:=& d(\frac{P_{m}}{\sqrt{W}}dX^{m}+\frac{P_{r}}{\sqrt{W}}dX^{r} -\overline{\Psi}\Gamma_{-}d\Psi)=0, \\
\phi_{2}:=& \oint_{C_{s}}(\frac{P_M}{\sqrt{W}}dX^M+\frac{P_{r}}{\sqrt{W}}dX^{r} -\overline{\Psi}\Gamma_{-}d\Psi)= 0,
   \end{aligned}
\end{equation}
associated with a residual symmetry of the theory: the infinite group of diffeomorphims preserving the Riemann basis $\Sigma$. So far, we have described the compactified supermembrane with no distinction between sectors. 
We now impose an extra topological
restriction on the winding maps \cite{mrt}: the irreducible winding constraint,
 \bea\label{central}
 \int_{\Sigma}dX^r\wedge dX^s=n\epsilon^{rs}Area(T^2)\quad n\ne 0, r,s=1,2.
 \eea
where $Area(T^2)=(2\pi R)^2 Im\tau$. \newline
-When this condition holds, $(n\ne 0)$ we refer to it as the supermembrane with central charge theory \cite{mor} and it  implies that the winding matrix $\mathbb{W}= \begin{pmatrix} l_{1}& l_{2}\\m_{1} & m_{2}\end{pmatrix}$ has  $det \mathbb{W}=n\ne 0$. Globally it corresponds to a sector, with nontrivial symplectic torus bundle with monodromy $\rho$ in $SL(2,Z)$, characterized by having $H^2(\Sigma,\mathbb{Z_{\rho}})\ne 0$ \cite{gmmpr},\cite{gmpr2}. \newline
-When this condition vanishes ($n=0$), this corresponds to a sector that we will call from now on compactified M2, $n=0$. Globally it corresponds to a trivial symplectic torus bundle over the base,  lets choose for symplicity the flat 2-torus $\Sigma_1$. It is characterized by having a trivial class of  $H^2(\Sigma_1,\mathbb{Z})$.

The Mass operator of the compactified supermembrane with winding and KK contribution \cite{sl2z}, \cite{schwarz}, is
\bea\label{28} Mass^{2}=T_{11}^{2}((2\pi R)^{2}n Im \tau)^{2}+
\frac{1}{R^{2}}((m_{1}^{2}+(\frac{m\vert q\tau-p\vert}{R Im\tau})+ T_{11}^{2/3}H\eea where the $H$ is defined in terms of the above hamiltonian $\mathcal{H}$ once the winding contribution has been extracted $H= \mathcal{H}-
T_{11}^{-2/3}\int_{\Sigma}\sqrt{W}\frac{T_{11}^{2}}{4}\{X^{r}_{h},X^{s}_{h}\}^{2}$ \cite{sl2z}. In the case of the sector of the  compactified M2, sector $n=0$, the winding contribution vanishes.

\section{T-duality in the Supermembrane Theory}
In this section we introduce the T-duality transformations for the supermembrane theory \cite{gmpr2}. This goes beyond the T-duality of superstring theory. The T-duality transformation we consider, is a nonlinear map which interchange the winding modes $\mathbb{W}$, associated to the cohomology of the base manifold with the KK charges, $Q =(p,q)$ associated to the homology of the target torus together with a transformation of the real moduli $R \to \frac{1}{R}$ and complex moduli $\tau \to \widetilde{\tau}$, both in a nontrivial way. In the following all transformed quantities under T-duality are denoted by a tilde.\newline
 In order to define the T-duality transformation we introduce the following \cite{sl2z}(47)  dimensionless variables
\bea
\mathcal{Z}:= T_{11}A\widetilde{Y}\quad \mathcal{\widetilde{Z}}:= T_{11} \widetilde{A}Y
\eea

where $T_{11}$ is the supermembrane tension, $A = (2\pi R)^2 Im\tau$ is the area of the target torus and $Y=\frac{R Im\tau}{\vert q\tau -p\vert}$. The tilde variables $\widetilde{A}, \widetilde{Y}$ are the transformed quantities under the T-duality. See (\ref{trans}) for the explicit value of $\mathcal{Z}$. The T-duality transformation we introduce is given by \cite{gmpr2}:
\bea\label{tt1}
\begin{aligned}
\textrm{The moduli}:& \quad \mathcal{Z}\widetilde{\mathcal{Z}}=1,\quad\widetilde{\tau}=\frac{\alpha \tau+\beta}{\gamma\tau +\alpha}; \\
\textrm{The charges}:& \begin{pmatrix}\widetilde{p}\\ \widetilde{q}\end{pmatrix}=\Lambda_0 \begin{pmatrix} p\\ q\end{pmatrix},
\begin{pmatrix}\widetilde{l}_1& \widetilde{l}_2\\ \widetilde{m}_1 & \widetilde{m}_2\end{pmatrix}=\Lambda_0^{-1}\begin{pmatrix}l_1^{'} & l_2^{'}\\ m_1^{'} & m_2^{'} \end{pmatrix}.
\end{aligned}
\eea
With $\Lambda_0= \begin{pmatrix}\alpha & \beta\\ \gamma & \alpha\end{pmatrix}\in SL(2,Z)$. In the above definition the T-dual supermembrane corresponds to a new supermembrane where the role of winding and KK charges interchanged, i.e. the KK modes are mapped onto the winding modes $
\begin{pmatrix}
\widetilde{p}\\ \widetilde{q}\end{pmatrix}=\begin{pmatrix}
l_1^{'}\\ m_1^{'}\end{pmatrix}$ and viceversa.
The above property together with the condition $Z\widetilde{Z}=1$ ensure that $(\textrm{T-duality})^2= \mathbb{I}$, the main property of T-duality.  The explicit transformations of the real modulus, obtained from the above T-duality transformation is
\begin{equation}\label{trans}
\widetilde{R}= \frac{\vert \gamma\tau+\alpha\vert \vert q\tau-p\vert ^{2/3}}{T_{11}^{2/3}(Im \tau)^{4/3}(2\pi)^{4/3}R},\qquad \\ \textrm{with}\quad\widetilde{\tau}=\frac{\alpha \tau+\beta}{\gamma\tau +\alpha} \quad \textrm{and}\quad \mathcal{Z}^3=\frac{T_{11} R^3 (Im\tau)^2}{\vert q\tau-p\vert}
\end{equation}
The winding modes and KK charge contribution in the mass squared formula transform in the following way:
\bea
\begin{aligned}
T_{11}n^2 A^2 &= \frac{n^2}{\widetilde{Y}^2}\mathcal{Z}^2,\qquad
\frac{m^2}{Y^2}&= T_{11}^2m^2 \widetilde{A}^2\mathcal{Z}^2.
\end{aligned}
\eea
To see how the $H_1$ (\ref{28}) transforms under T-duality it is important to realize the transformation rules for the fields,
 \bea
 \begin{aligned}
 dX^m=u d\widetilde{X}^m,\quad
 d\widetilde{X}=u e^{i\varphi}dX,\quad
 A=u e^{i\varphi}\widetilde{A}\quad
 \Psi=u^{3/2} \widetilde{\Psi},\quad
 \overline\Psi=u^{3/2}\widetilde{\overline\Psi}
 \end{aligned}
 \eea
 Where $u=\mathcal{Z}^2=\frac{R\vert \gamma\tau+\alpha\vert}{\widetilde{R}}$, $\varphi$ a phase defined in (3.22) of \cite{gmpr2} and $dX=dX^1+idX^2$ and respectively, its dual $d\widetilde{X}$ is\bea
\begin{aligned}
d\widetilde{X}=2\pi \widetilde{R}[(\widetilde{m}_1\widetilde{\tau}+\widetilde{l_1})d\widehat{X}^1+(\widetilde{m}_2\widetilde{\tau}+\widetilde{l}_2)d\widehat{X}^2]
\end{aligned}
\eea
The phase $e^{i\varphi}$ cancels with the h.c. the transformation of the Hamiltonian.
 The relation between the hamiltonians through a T-dual transformation is
 \bea
 H=\frac{1}{\widetilde{\mathcal{Z}}^{8}}\widetilde{H},\quad \widetilde{H}= \frac{1}{\mathcal{Z}^8}H.
 \eea

 We thus obtain for the mass squared formula the following identity,
 \bea
 M^2 = T_{11}^2 n^2 A^2 + \frac{m^2}{Y^2}+ T_{11}^{2/3}H =\frac{1}{\widetilde{\mathcal{Z}}^2}(\frac{n^2}{\widetilde{Y}^2}+ T_{11}^2 m^2 \widetilde{A}^2)+ \frac{T_{11}^{2/3}}{\widetilde {\mathcal{Z}}^8}\widetilde{H}.
\eea
%

\subsection{T-Duality on Symplectic Bundles}
T-duality does not only acts at local level but also globally. 
 As shown in \cite{gmpr2} one can define an equivalence class with the elements of the coinvariant group associated to the monodromy group $G$, 
$
 \{\mathcal{Q}+g\widehat{\mathcal{Q}}-\widehat{\mathcal{Q}}\},
$
such that for any $g\in G$ and $\widehat{\mathcal{Q}}\in H_1 (T^2)$, it characterizes one symplectic torus bundle. In the formulation of the supermembrane on that geometrical structure, $\mathcal{Q}$ are identified with the KK charges. The action of $G$, the monodromy group, leaves the equivalence class invariant. We now consider the duality transformation introduced previously. Under the duality transformation the equivalence class transform as

\bea
\{\mathcal{Q}+g\widehat{\mathcal{Q}}-\widehat{\mathcal{Q}}\} \to \{\Lambda_0\mathcal{Q}+(\Lambda_0 g \Lambda_0^{-1})\Lambda_0\widehat{\mathcal{Q}}-\Lambda_0\widehat{\mathcal{Q}}\},
 \eea
 hence for the dual bundle it holds,
$
 \{\Lambda_0\begin{pmatrix}l_1\\ m_1\end{pmatrix}+(\Lambda_0 g \Lambda_0^{-1})\begin{pmatrix}\widehat{l_1}\\ \widehat{m_1}\end{pmatrix}-\begin{pmatrix}\widehat{l_1}\\ \widehat{m_1}\end{pmatrix}\},$
 That is, as an element of the coinvariant group of $\Lambda_0 G \Lambda_0^{-1}$. We then conclude that the duality transformation, in addition to the transformation on the moduli $R,\tau$, also maps the geometrical structure onto an equivalent symplectic torus bundle with monodromy $\Lambda_0 G\Lambda_0^{-1}$. We notice that the transformation depends crucially on the original equivalence class of the coinvariant group. So for a nonequivalent symplectic torus bundle the dual transformations is realized with a different $SL(2,Z)$ matrix $\Lambda_0$. Now we are in position to determine the T-duality as a natural symmetry for the family of
supermembranes with central charges. We take: \bea\label{la} \widetilde{Z}=Z=1\Rightarrow
T_0=\frac{\vert q\tau-p\vert}{R^3 (Im\tau)^2}. \eea
It imposes a relation between the energy scale of the tension of the supermembrane and the moduli of
the torus fiber and that of its dual. Indeed we can think in two different ways: given the values of the moduli it fixes the allowed tension $T_0$ or on the other way around, for a fixed tension $T_0$, the radius, the Teichmuller parameter of the 2-torus, and the KK charges satisfy (\ref{la}).
\section{String Theory Limit}
We then
consider within the physical configurations of the supermembrane with central charges, the
string-like configurations \bea X^{m}=X^{m}(\tau,
q_{1}\widehat{X}^{1}+q_{2}\widehat{X}^{2}),\quad  A^{r}=A^{r}(\tau,
q_{1}\widehat{X}^{1}+q_{2}\widehat{X}^{2}) ,\eea where $q_{1},q_{2}$
are relative prime integral numbers. $X^{m}, A^{r}$ are scalar
fields on the torus $\Sigma$, a compact Riemann surface, hence they
may always be expanded on a Fourier basis in term of a double
periodic variable of that form. The restriction of $q_{1}, q_{2}$ to
be relatively prime integral numbers arises from the global
periodicity condition. On that configurations all the brackets \bea
\{X^{m},X^{n}\}=\{X^{m}, A^{r}\}=\{A^{r},A^{s}\}=0 \eea vanish.
We then obtain the final expression for the mass
contribution of the string states \cite{sl2z}:
 \bea\label{78} M_{11}^{2}\vert_{SC}= (n
T_{11}A)^2+(\frac{m}{Y})^{2}+ 8\pi^{2} R_{11}T_{11}\vert q\tau -p \vert
(N_{L}+N_{R}) \eea where $(p,q)$ are relatively prime. We notice that $(p,q)$ may be interpreted as
the wrapping of the membrane around the two cycles of the target torus.
 The corresponding change in the harmonic sector is \cite{sl2z}
\bea
dX_{h}=(qmd\widetilde{X}^{1}+pd\widetilde{X}^{2})
+\widetilde{\tau}(-Q n d\widetilde{X}^{1}+Pd\widetilde{X}^{2}),
\eea
the hamiltonian is invariant under that change. $p,q$ and $Q,P$ are now the
winding numbers of the supermembrane. Given $p,q$ there always exist
$Q$ and $P$
with the above property, although the correspondence is not unique. The $(p,q)$
type IIB strings may indeed be interpreted as different wrappings of the supermembrane with central charges.
This nice interpretation was first given in \cite{schwarz}.

The $(p,q)$ $IIB$ string compactified on a circle of radius $R_{B}$
has tension \cite{schwarz} \bea T_{(p,q)}^{2}=\frac{\vert
q\lambda_{0}-p  \vert^{2}}{Im\lambda_{0}}T^{2} \eea  where $T= T_{11}^{2/3}$ is the string tension and
$\lambda_{0}=\xi_{0}+ie^{-i\phi_{0}}$ with $\xi$ and $\phi$ 
identified with the scalar fields of the type IIB theory, $\phi$
corresponds to the dilaton fields. $\lambda_{0}$ is the asymptotic
value of $\lambda$ -the axion-dilaton of the type IIB theory-
specifying the vacuum of the theory. The perturbative spectrum of
the $(p,q)$
 IIB string is \cite{schwarz},
\bea\label{82} M_{B}^{2}=(\frac{n}{R_{B}})^{2}+(2\pi R_{B}m
T_{(p,q)})^{2}+4\pi T_{(p,q)} (N_{L}+N_{R}) .\eea If we use following
\cite{schwarz} a factor $\beta^{2}$ to identify term by term
 both mass formulas ($M_{11}=\beta M_B$), since there were obtained using different metrics, one gets
\bea\label{83} \tau= \lambda_{0},\quad  \beta^{2}=
 \frac{T_{11}A_{11}^{1/2}}{T},\quad R_{B}^{-2}= T T_{11}
A_{11}^{3/2}. \eea  
They were obtained by counting modes under some
assumptions on the supermembrane wrapping modes, as mentioned on one
of the footnotes \cite{schwarz}. Here we have derived the
expressions from a consistent definition of the supermembrane with central charges. \newline 
The identification of (\ref{78}) to the mass formula of IIA string
compactified on a circle of radius $R_{A}$ and tension $T_{A}$ may
also be performed. In order to have a consistent identification one
has to take $Re\tau=0$, $p=1$ and hence $q=0$ in (\ref{78}). The
mass formula for the perturbative spectrum of type IIA is \bea
M_{A}^{2}=(\frac{m}{R_{A}})^{2}+(2\pi R_{A}n T_{A})^{2}+4\pi T_{A}
(N_{L}+N_{R}) \eea Identification after the limit process of the winding contributions and KK ones
 using a factor $(\beta\gamma)$ to compare the $mass^{2}$ formulas,
since they are obtained using different metrics, yields
$
 R_{A}=\beta\gamma R_{11},\quad
T_{A}=\gamma^{-2}(Im\tau)^{1/2} T
$
which imply
\bea
(2\pi R_{A}R_{B})=(\frac{1}{T_{A}T_{(p,q)}})^{1/2}
\eea
We have thus obtained the $(p,q)$ IIB and IIA perturbative spectrum,
when compactified on circles $R_{B}$ and $R_{A}$ respectively, from the string states on
the supermembrane with central charges. \newline
In this limit, by restricting the worldvolume configurations of the M2 to those of the string  \cite{sl2z}, we exactly recover the mass operator of the IIB theory as was formerly conjectured by Schwarz. The gain is that from the supermembrane with central charges the pure membrane excitations are known.  If now a T-duality is performed on the supermembrane M2 mass operator restricted to string-like configurations, then an SL(2,Z) non-perturbative multiplet of IIA is obtained \cite{sl2z}.
\paragraph{String T-duality transformation limit.}
 Now we can take directly the limit of the T-duality transformations of the supermembrane to recover the standard T-duality transformations for the closed string operator. 
Before taking the limit it is convenient to consider a redefinition of $X^1$ and $X^2$. The coordinates that wrap on the $T^2$. We take 
$
X^1\to\frac{X^1}{T_{11}^{1/6}R_{11}^{1/2}}, X^2\to T_{11}^{1/6}R_{11}^{1/2} X^2,
$
such that the Lie brackets $\{X^1,X^2\}^2$ will remain invariant under the redefinition. Now we consider the following limit for the torus collapsing into a circle, by imposing $ R_{11}\to 0$ . Since we want to ensure $A_{11}Y$ remains finite but $A_{11}\to 0$, necessarily $q=0$, for arbitrary $p$. Indeed this is equivalent to consider the KK charges $(p,q)=m ( 1,0)$ for $m=p$. By using the M2 T-duality transformation it can also be seen that $\tilde{A}_{11}\to 0$, so the dual also corresponds to a string. Moreover substituting in the previous definitions it can be seen  $R_A$ is finite implies $R_B$ finite.
We will define  the following for the torus degenerating into a circle $S^1$ 
\bea
R_1=\frac{R_{11}^{1/2}}{T_{11}^{1/6}}; \quad R_{2}=R_{11}^{3/2}T_{11}^{1/6} Im\tau
\eea

Since $R_{11}\to 0$ , it implies $R_1\to 0$ but $R_2$ is finite, so it corresponds to a closed curve that topologically is a circle. Now we re-express the winding condition in terms of the new variables. In terms of the new variables we get
\bea
\oint_{\mathcal{C}_s} dX^1= 2\pi R_1 l_s;\quad \oint_{\mathcal{C}_s} dX^2= 2\pi R_2 m_s,
\eea
Since $R_1\to 0$, although $l_s$ is taking finite, the first winding condition vanishes and the only residual winding condition is associated to the $S^1$ modulus is $R_2$. The former T-duality relations of the moduli in this limit become reduced to:
\bea
  Z\widetilde{Z}=1 \vert_{string},\quad \Rightarrow\quad
   T_{M2}^{4/3} R_2^3\widetilde{R}_2^3=1\to \widetilde{R}_2=\frac{\alpha^{'}}{R_2}. \eea
 where $\widetilde{R}_2= T_{11}^{1/2}\widetilde{A}^{1/2}\widetilde{Y}^{1/2}$. This defines for $T_{(p,q)}=T$ on the IIB string side precisely the duality relation of the strings.
The transformation on the charges and windings are given by (\ref{tt1}) and we finally obtain:\bea \{R;(l_1,m)\}\stackrel{T-duality}{\longrightarrow}\{\widetilde{R}=\frac{\alpha^{'}}{R}; (m,l_1)\}\eea where $m$ is the common factor between the charges.  
\section{Discussion and Conclusions}
We showed the existence of a new $Z_2$ symmetry that plays the role of T-duality in M-theory interchanging the winding and KK charges but leaving the hamiltonian invariant. The supermembrane compactified on a torus realizes this duality as an exact symmetry of the theory in both sectors ($n =0$, $n\ne 0$). This is a relevant property expected for a sector of M-theory. When only string-like states are considered but it is performed this generalized T-duality of the supermembrane we obtain the mass operator IIB and the corresponding dual realizes the type IIA with all nonperturbative $SL(2,Z)$ multiplet. If we take the limit of the T-duality transformation of the supermembrane into the standard one, then only the standard $(p,q)= (1,0)$ mass type IIA operator is allowed meanwhile IIB mass operator is unchanged.
\section{Acknowledgements}Part of the work of MPGM was funded by the Spanish
Ministerio de Ciencia e Innovaci\'on (FPA2006-09199) and the
Consolider-Ingenio 2010 Programme CPAN (CSD2007-00042). The work of A. R. is funded by Proyecto FONDECYT 1121103.

\end{document}